\documentclass[aps,pre,onecolumn,superscriptaddress,showpacs]{revtex4}

\usepackage{amssymb}
\usepackage{epsfig}
\usepackage{graphicx}
\usepackage{float}
\usepackage{amsmath}
\usepackage{times}
\usepackage{multirow}

\begin{document}

\title{Percolation on interacting networks with feedback-dependency links}
\author{Gaogao Dong}\email{dfocus.gao@gmail.com}
\affiliation{Nonlinear Scientific Research Center, Faculty of
Science, Jiangsu University, Zhenjiang, 212013, China}

\affiliation{Center for Polymer Studies and Department of Physics,
Boston University, Boston, MA 02215, USA}

\author{Lixin Tian}
\affiliation{Nonlinear Scientific Research Center, Faculty of
Science, Jiangsu University, Zhenjiang, 212013, China}

\author{Ruijin Du}
\affiliation{Nonlinear Scientific Research Center, Faculty of
Science, Jiangsu University, Zhenjiang, 212013, China}

\affiliation{College of Mathematics Science, Chongqing Normal
University, Chongqing, 401331, China}

\affiliation{Center for Polymer Studies and Department of Physics,
Boston University, Boston, MA 02215, USA}

\author{Min Fu}
\affiliation{Nonlinear Scientific Research Center, Faculty of
Science, Jiangsu University, Zhenjiang, 212013, China}

\author{H. Eugene Stanley}
\affiliation{Center for Polymer Studies and Department of Physics,
Boston University, Boston, MA 02215, USA}

\begin{abstract}
When real networks are considered, coupled networks with
connectivity and feedback-dependency links are not rare but more
general. Here we develop a mathematical framework and study
numerically and analytically percolation of interacting networks
with feedback-dependency links. We find that when nodes of between
networks are lowly connected, the system undergoes from second order
transition through hybrid order transition to first order transition
as coupling strength increases. And, as average degree of each
inter-network increases, first order region becomes smaller and
 second-order region becomes larger but hybrid order region almost keep constant.
 Especially, the results implies that
average degree $\bar{k}$ between intra-networks has a little
influence on robustness of system for weak coupling strength, but
for strong coupling strength corresponding to first order transition
system become robust as $\bar{k}$ increases. However, when average
degree $k$ of inter-network is increased, the system become robust
for all coupling strength. Additionally, when nodes of between
networks are highly connected, the hybrid order region disappears
and the system first order region becomes larger and
 second-order region becomes smaller. Moreover, we find that the existence
of feedback dependency links between interconnecting networks makes
the system extremely vulnerable by comparing non-feedback condition
for the same parameters.
\end{abstract}

\maketitle

\section{Introduction}
Complex networks have been studied extensively owing to their
relevance to many real systems, where nodes of the network can be
grouped by connectivity links. During the past decade, complex
theory is exclusively focused on the single and isolated networks
\cite{Watts1998,Bar1999,Albert2002,Cohen2000,Callaway2000,Dorogovtsev2003,Satorras2006,A.Bashan2012,Song2005,Havlin2010,Caldarelli2007,Newman2010,Hu2011_1,Liu2012,Rong2010,Yang2010,Dai2013,Li2011,Li20112,Liuy2012}
. In reality, networks rarely appear in isolation, where have wide
variety of coupled networks. Recently, there has been a turning
point in accordance with the advent of concepts of interdependent
networks and interacting networks
\cite{Havlin22010,Buldyrev2010,Leicht2010,Bashan2011,Parshani2010,Dong2012,Gao20111,Huang2011,Liw2012,Parshani2011,DongGG2013,Zhou2013,Shao2009,DongGGEPL2013,Hu2011,Gao20113,Wang2013}.
Buldyrev et al. developed a framework for understanding the
robustness of couple networks with only dependency links between
nodes of two networks, which subject to cascading failures according
to Italy blackout on 2003. Their findings suggest that dependency
links between nodes of two networks have an important influence on
designing resilient infrastructures \cite{Buldyrev2010}. Meanwhile,
Leicht et al. developed a mathematical framework based on generating
functions for analyzing a system of $n$ coupled networks with only
connectivity links between nodes of two networks. Their findings
highlight the extreme lowering of the percolation threshold possible
once connectivity links between networks are taken into account
\cite{Leicht2010}. Moreover, Shao et al. investigated cascading
failures of coupled networks with multiple support-dependence
relations by considering unidirectional support dependency links
between nodes of two networks. Their model can help to further
understand real-life coupled network systems, where complex
dependence-support relations exists \cite{Shao2009}. In fact, real
network often contain both types of links, dependency and
connectivity links \cite{Parshani2011,Hu2011,Bashan2011}. Parshani
et al. modeled single networks with two different links and
discussed it's robustness. They found that networks with high
density of dependency links are extremely vulnerable, but networks
with a low density of dependency links are significantly more robust
\cite{Parshani2011}. Hu et al. studied coupled networks with both
connectivity and dependency links between nodes of two networks,
where dependency links is no feedback condition. Their findings
conclude that the connectivity links increase the robustness of the
system, while the interdependency links decrease its robustness
\cite{Hu2011}. Gao et al. researched the robustness of $n$ coupled
loop networks with the condition of feedback dependency links
between nodes of two networks. They pointed out that coupled
networks is extremely vulnerable as feedback dependency links exist
between two networks \cite{Gao20113}. When real networks are taken
into account, coupled networks with feedback-dependency and
connectivity links are not rare but more general. Here we develop a
mathematical framework to study the robustness of two interacting
networks with feedback-dependency links.

\section{Framework}
\begin{figure}[H]
\centering \scalebox{0.2}[0.2]{\includegraphics{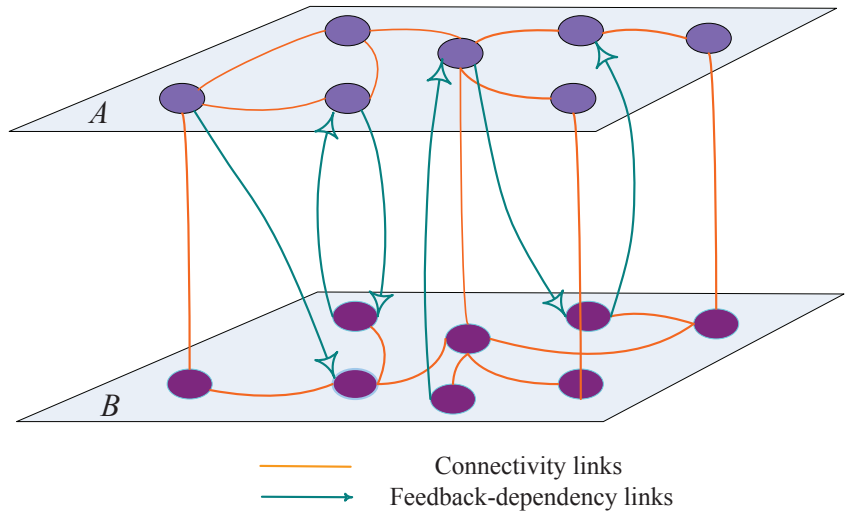}}
\caption{Demonstration of interacting networks with feedback
dependency links. The feedback dependency links between network $A$
and network $B$ are random and directional. The nodes of $A$ and $B$
are randomly connected with connectivity links.}
\end{figure}
For two networks $A$ and $B$ of sizes $N_{A}$ and $N_{B}$, we assume
that they are coupled by both dependency and connectivity links. For
the case of dependency links, the two networks are partially
coupled, which means dependency links between two fractions $q_{A}$
and $q_{B}$ of nodes in $A$ and $B$ networks satisfies the feedback
condition (as shown in Fig. 1). For the other case, connectivity
links connecting nodes within each network and between the networks,
which can be presented by degree distributions
$P^{A}_{(k_{A},k_{AB})}$, $P^{B}_{(k_{B},k_{BA})}$ respectively,
where $P^{A}_{(k_{A},k_{AB})}$ and $P^{B}_{(k_{B},k_{BA})}$ denote
the probability of an node in $A$ or ($B$) to have $k_{A}$ or
($k_{B}$) links to nodes in the same network and $k_{AB}$ or
($k_{BA}$) links toward other network. When nodes fail in a network,
all connectivity links connected to these nodes fails, causing other
nodes to disconnect from the network. Since dependency relations
between networks, interdependent nodes in other network also remove
along with their connectivity links. We assume that a functional
node in network $A$ ($B$) must belong to the giant component of
network $A$ ($B$). When this cascading process occur, it will stop
if nodes that fail in one step do not cause additional failures and
stabilizes with giant component.

When a fraction $1-p$ of $A$ nodes are initially removed,
$g_{A}(\omega_{t},\varpi_{t})$ and $g_{B}(\omega_{t},\varpi_{t})$
 are equal to the fraction of nodes in the giant components of
 networks $A$ and $B$ at step $t$, after removal of fractions
 $1-\omega_{t}$ and $1-\varpi_{t}$, respectively. Thus, the
 cascading dynamics can be described by

\begin{equation}
\begin{split}
& \omega_{1}=p, \varpi_{1}=1, P_{1}^{A}=\omega_{1}g_{A}(\omega_{1},\varpi_{1}),\\
& \varpi_{2}=1-q_{B}(1-P_{1}^{A}),
P_{2}^{B}=\varpi_{2}G_{B}(\omega_{1},\varpi_{2}),\\
& \omega_{2}=p[1-q_{A}(1-P_{2}^{B})],P_{2}^{A}=\omega_{2}g_{A}(\omega_{2},\varpi_{2}),\\
& \varpi_{3}=1-q_{B}(1-P_{2}^{A}),P_{3}^{B}=\varpi_{3}g_{B}(\omega_{2},\varpi_{3}),\\
& \omega_{3}=p[1-q_{A}(1-P_{3}^{B})],P_{3}^{A}=\omega_{3}g_{A}(\omega_{3},\varpi_{3}),\\
& \cdots\\
& \varpi_{t}=1-q_{B}(1-P_{t-1}^{A}),P_{t}^{B}=\varpi_{t}g_{B}(\omega_{t-1},\varpi_{t}),\\
& \omega_{t}=p[1-q_{A}(1-P_{t}^{B})],P_{t}^{A}=\omega_{t}g_{A}(\omega_{t},\varpi_{t}).\\
\end{split}
\end{equation}

Where, $P_{t}^{A}$ ($P_{t}^{B}$) is the corresponding giant
components of network $A$ ($B$).

For $\omega_{t}$, $\varpi_{t}$, $P_{t}^{B}$ and $P_{t}^{A}$, at
$t\rightarrow \infty$, since eventually the clusters stop
fragmenting. Thus, at steady state, the expression of system can be
given by
\begin{equation}
\begin{split}
& \varpi_{\infty}=1-q_{B}(1-P_{\infty}^{A}),P_{\infty}^{B}=\varpi_{\infty}g_{B}(\omega_{\infty},\varpi_{\infty}),\\
&
\omega_{\infty}=p[1-q_{A}(1-P_{\infty}^{B})],P_{\infty}^{A}=\omega_{\infty}g_{A}(\omega_{\infty},\varpi_{\infty}).
\end{split}
\end{equation}

\section{Theory}
In this paper, we consider the case where all degree distributions
of the connectivity intra- and interlinks are Poissonian. Thus, the
two-dimensional generating function are as follows \cite{Leicht2010,Hu2011}\\
\begin{equation}
\begin{split}
& G_{0}^{A}(x_A,x_B)=\sum_{k_{A},\bar{k}_{A}}P_{k_{A},\bar{k}_{A}}^{A}x_{A}^{k_{A}}x_{B}^{\bar{k}_{A}},\\
& G_{0}^{B}(x_A,x_B)=\sum_{k_{B},\bar{k}_{B}}P_{k_{B},\bar{k}_{B}}^{B}x_{A}^{\bar{k}_{B}}x_{B}^{k_{B}},\\
&
G_{1}^{AB}(x_A,x_B)=\sum_{k_{A},\bar{k}_{A}}\frac{(\bar{k}_{A}+1)P_{k_{A},\bar{k}_{A}+1}^{A}}{\displaystyle
\sum_{k^{'}_{A},\bar{k}^{'}_{A}}\bar{k}^{'}_{A}P_{k^{'}_{A},\bar{k}^{'}_{A}}^{A}}x_{A}^{k_{A}}x_{B}^{\bar{k}_{A}}.\\
\end{split}
\end{equation}

where, $(\bar{k}_{A}+1)P_{k_{A},\bar{k}_{A}+1}^{A}$ is the
probability of following a randomly chosen $\bar{k}_{A}$ link
connecting an $A$ node of degree $k_{A}$ to a $B$ node with excess
$\bar{k}_{A}$ degree and $G_{1}^{AB}(x_A,x_B)$ is generating
function of this distribution. Accordingly, the other three excess
generating functions, $G_{1}^{AA}, G_{1}^{BA}, G_{1}^{BB}$, can be
obtained \cite{Leicht2010,Hu2011}
\begin{equation}
\begin{split}
& G_{0}^{AA}(x_{A})=e^{k_{A}(x_{A}-1)},\\
& G_{0}^{AB}(x_{B})=e^{\bar{k}_{A}(x_{B}-1)},\\
& G_{0}^{BA}(x_{A})=e^{\bar{k}_{B}(x_{A}-1)},\\
& G_{0}^{BB}(x_{B})=e^{k_{B}(x_{B}-1)}.
\end{split}
\end{equation}

Thus, from Eqs.(3) and (4), the four excess function can be
presented
\begin{equation}
\begin{split}
& G_{1}^{AA}(x_{A}, x_{B})=G_{1}^{AB}(x_{A}, x_{B})=G_{0}^{A}(x_{A}, x_{B})=G_{0}^{AA}(x_{A})G_{0}^{AB}(x_{B})=e^{k_{A}(x_{A}-1)}e^{\bar{k}_{A}(x_{B}-1)},\\
& G_{1}^{BB}(x_{A}, x_{B})=G_{1}^{BA}(x_{A}, x_{B})=G_{0}^{B}(x_{A},
x_{B})=G_{0}^{BA}(x_{A})G_{0}^{BB}(x_{B})=e^{\bar{k}_{B}(x_{A}-1)}e^{k_{B}(x_{B}-1)}.
\end{split}
\end{equation}

After removal of $1-\omega$ and $1-\varpi$ fractions of network $A$
and $B$, from Eqs. (4) and (5), we have
\begin{equation}
\begin{split}
& g_{A}(\omega, \varpi)=1-G_{0}^{A}[1-\omega(1-f_{A}),1-\varpi(1-f_{BA})],\\
& g_{B}(\omega,
\varpi)=1-G_{0}^{B}[1-\omega(1-f_{AB}),1-\varpi(1-f_{B})].
\end{split}
\end{equation}

where,
\begin{equation}
\begin{split}
& f_{A}=G_{1}^{AA}[1-\omega(1-f_{A}),1-\varpi(1-f_{BA})],\\
& f_{B}=G_{1}^{AB}[1-\omega(1-f_{AB}), 1-\varpi(1-f_{B})],\\
& f_{AB}=G_{1}^{AB}[1-\omega(1-f_{A}),1-\varpi(1-f_{BA})],\\
& f_{BA}=G_{1}^{BA}[1-\omega(1-f_{AB}),1-\varpi(1-f_{B})].
\end{split}
\end{equation}

For cascading process, we compare our theoretical results obtained
from Eqs. (1), (4), (5), (6) and (7) with results of numerical
simulations as shown in Fig. 2. One can see that the simulation
results show excellent agreement with the theory.
\begin{figure}[H]
\centering \scalebox{0.5}[0.5]{\includegraphics{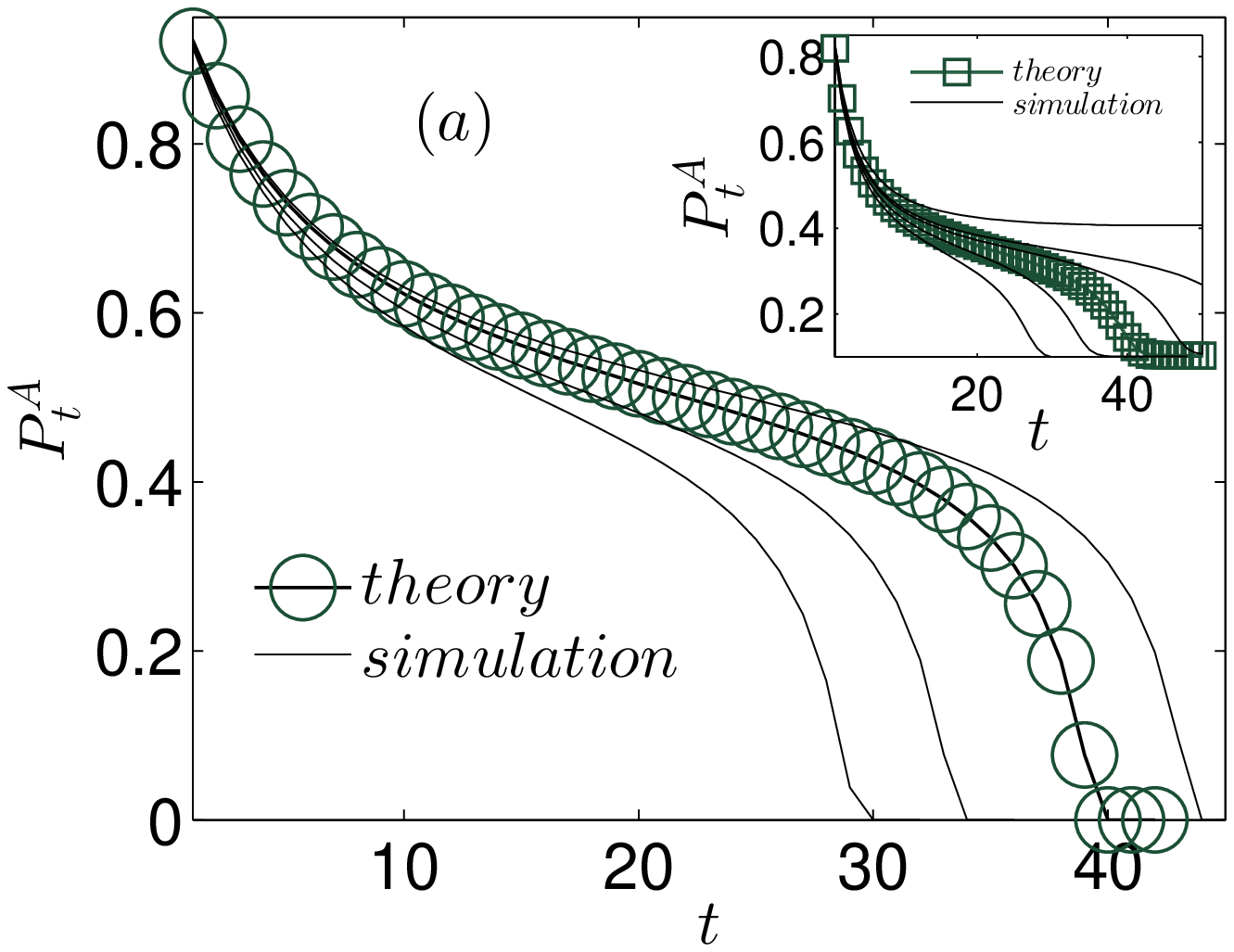}}
\centering \scalebox{0.5}[0.5]{\includegraphics{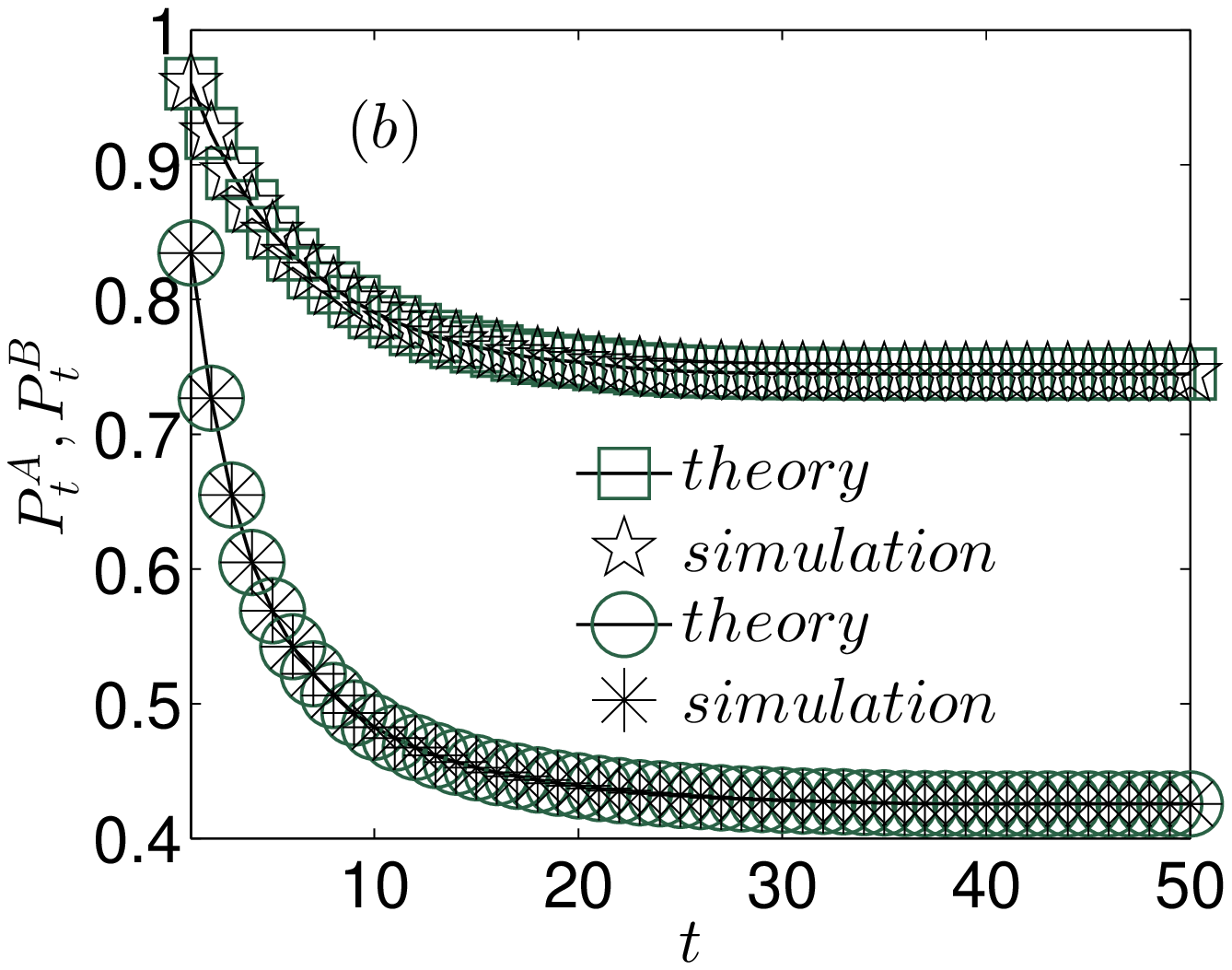}}
\caption{(a) Comparison between simulations and theory, the fraction
$P^{A}_{t}$ of giant component of network $A$ as a function of stage
$t$ with parameters $k_{A}=k_{B}=5$, $\bar{k}_{A}=\bar{k}_{B}=0.5$,
$q_B=1$ and $N^{A}=N^{B}=10^{5}$. We choose parameters $q_{A}=0.8$,
$p=0.928$ for main figure and $q_{A}=0.7$, $p=0.828$ for sub-figure.
(b) The fraction $P^{A}_{t}$ ($\bigcirc$) , $P^{B}_{t}$ ($\Box$) of
giant component of network $A$, $B$ as function of stage $t$ with
the same parameters as in (a) but $q_{A}=0.7$, $p=0.843$. The
simulation results are averaged over 50 realizations.}
\end{figure}
Submitting Eqs. (5), (6) and (7) into Eq. (2), at steady state, the
corresponding $P_{\infty}^{A}$ and $P_{\infty}^{B}$ are expressed
\begin{equation}
\begin{aligned}
P_{\infty}^{A}&=p[1-q_{A}(1-P_{\infty}^{B})][1-e^{-(k_{A}P_{\infty}^{A}+\bar{k}_{A}P_{\infty}^{B})}],\\
P_{\infty}^{B}&=1-q_{B}(1-P_{\infty}^{A})[1-e^{-(\bar{k}_{B}P_{\infty}^{A}+k_{B}P_{\infty}^{B})}].
\end{aligned}
\end{equation}

We presents comparison the theoretical predictions and simulations
for the giant components as a function of $q_{A}$ and $\bar{k}$ as
shown in Fig. 3(a)-(b).  One can see that the theory predictions
from Eq. (8) agrees well with simulation results for different set
of $q_{A}$ and $\bar{k}$ as shown in Fig. 3(a)-(b). Furthermore, we
can clearly find that as coupling strength $q_{A}$ increases, the
system undergoes second order transition to first order transition
through hybrid order transition, which means the size of the giant
component jumps at $p^{h,I}_{c}$ from a large value to a small value
then continuously decreases at $p^{h,II}_{c}$ to zero. And, Fig.
3(a)-(b) also presents corresponding critical fraction $p_{c}$,
including first and second order transition points $p^{I}_{c}$,
$p^{II}_{c}$, two hybrid order transition points $p^{h,I}_{c}$,
$p^{h,II}_{c}$. Additionally, the number of iterative failures
($NOI$) as a function of $\bar{k}$ and $p$ is shown in Fig. 3(c),
one can observe that $NOI$ has a peak at jump points, $p^{I}_{c}$
and $p^{h,I}_{c}$. Thus, it provides a useful and precise method for
identifying the transition points $p^{I}_{c}$ and $p^{h,I}_{c}$ by
computing $NOI$ as a function of $p$.
\begin{figure}[H]
\centering \scalebox{0.35}[0.35]{\includegraphics{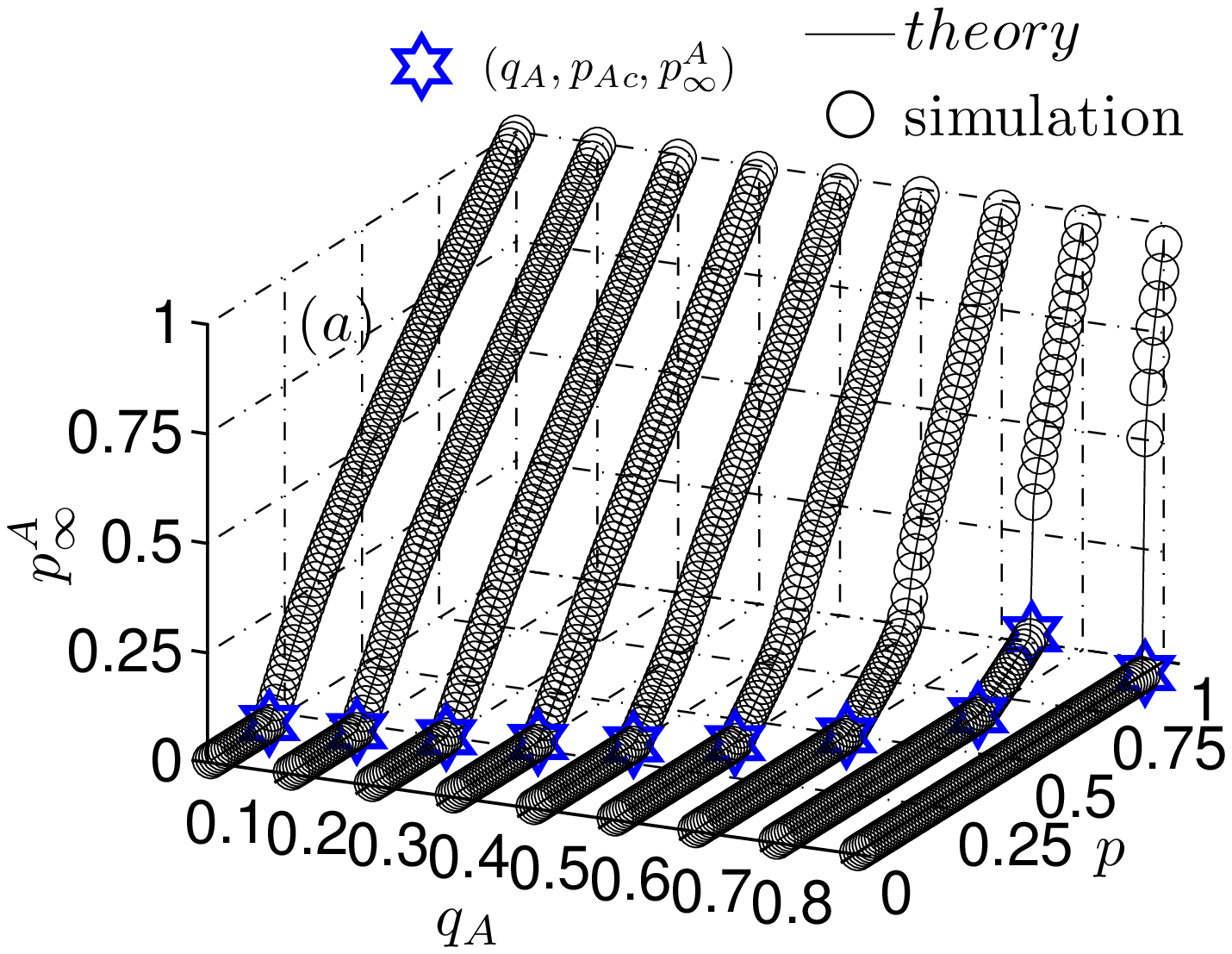}}
\centering \scalebox{0.35}[0.35]{\includegraphics{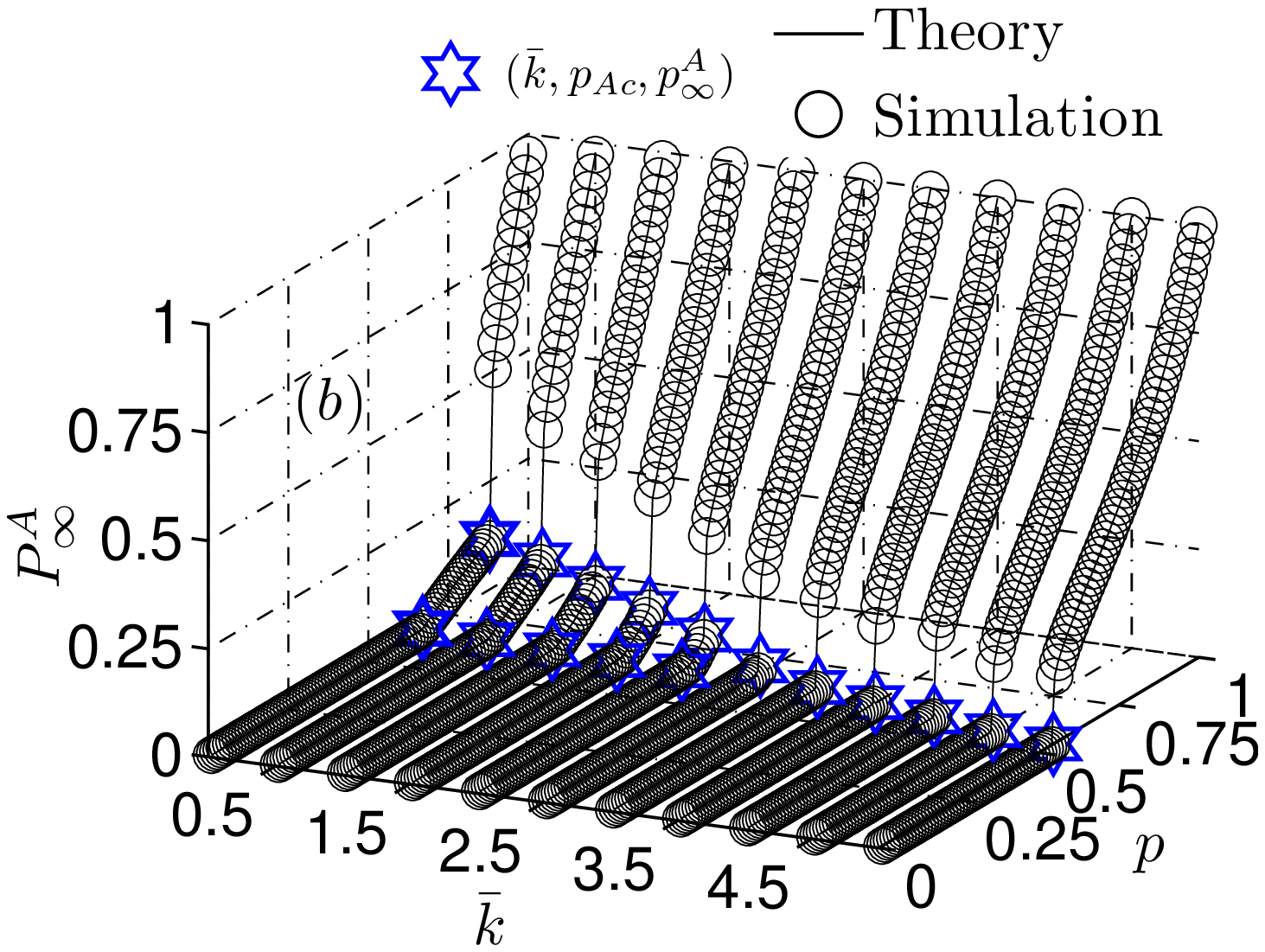}}
\centering \scalebox{0.35}[0.35]{\includegraphics{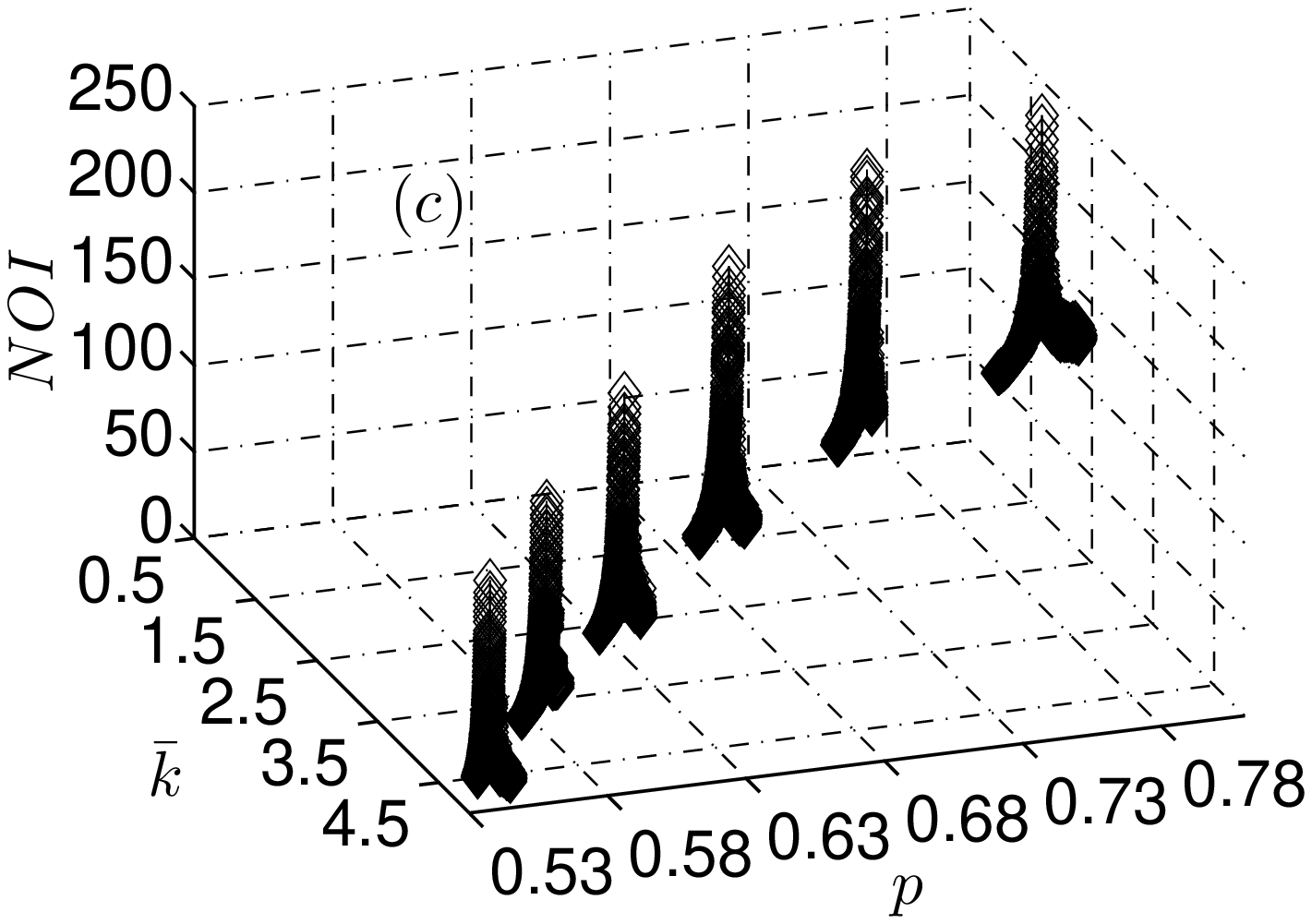}}
\caption{(a) Comparison between simulations and theoretical
predictions, the fraction of giant component $p^{A}_{\infty}$ as a
function of $p$ and $q_{A}$ with parameters $k_{A}=k_{B}=5$,
$\bar{k}_{A}=\bar{k}_{B}=0.5$, $N_{A}=N_{B}=10^{5}$. (b)
$p^{A}_{\infty}$ as a function of $p$ and $\bar{k}$ with the same
parameters as in (a) but $k_{A}=k_{B}=5$. (c) $NOI$ as a function of
$p$ and $\bar{k}$ with same parameter as in (b) from numerical
analysis. The simulation results are averaged over 50 realizations.}
\end{figure}
In fact, Eqs. (8) can be solved graphically as shown in Fig. 4. For
given parameters, Fig. 4 implies that the critical point $p^{I}_c$
and $p^{h,I}_c$ is the intersection of the two curves
$P^{A}_{\infty}(P^{B}_{\infty})$ and
$P^{B}_{\infty}(P^{A}_{\infty})$. Thus, the corresponding critical
manifold can be found from the tangential condition
\begin{equation}
\frac{dP^{A}_{\infty}}{dP^{B}_{\infty}}\frac{dP^{B}_{\infty}}{dP^{A}_{\infty}}=1
\end{equation}
\begin{figure}[H]
\centering \scalebox{0.35}[0.35]{\includegraphics{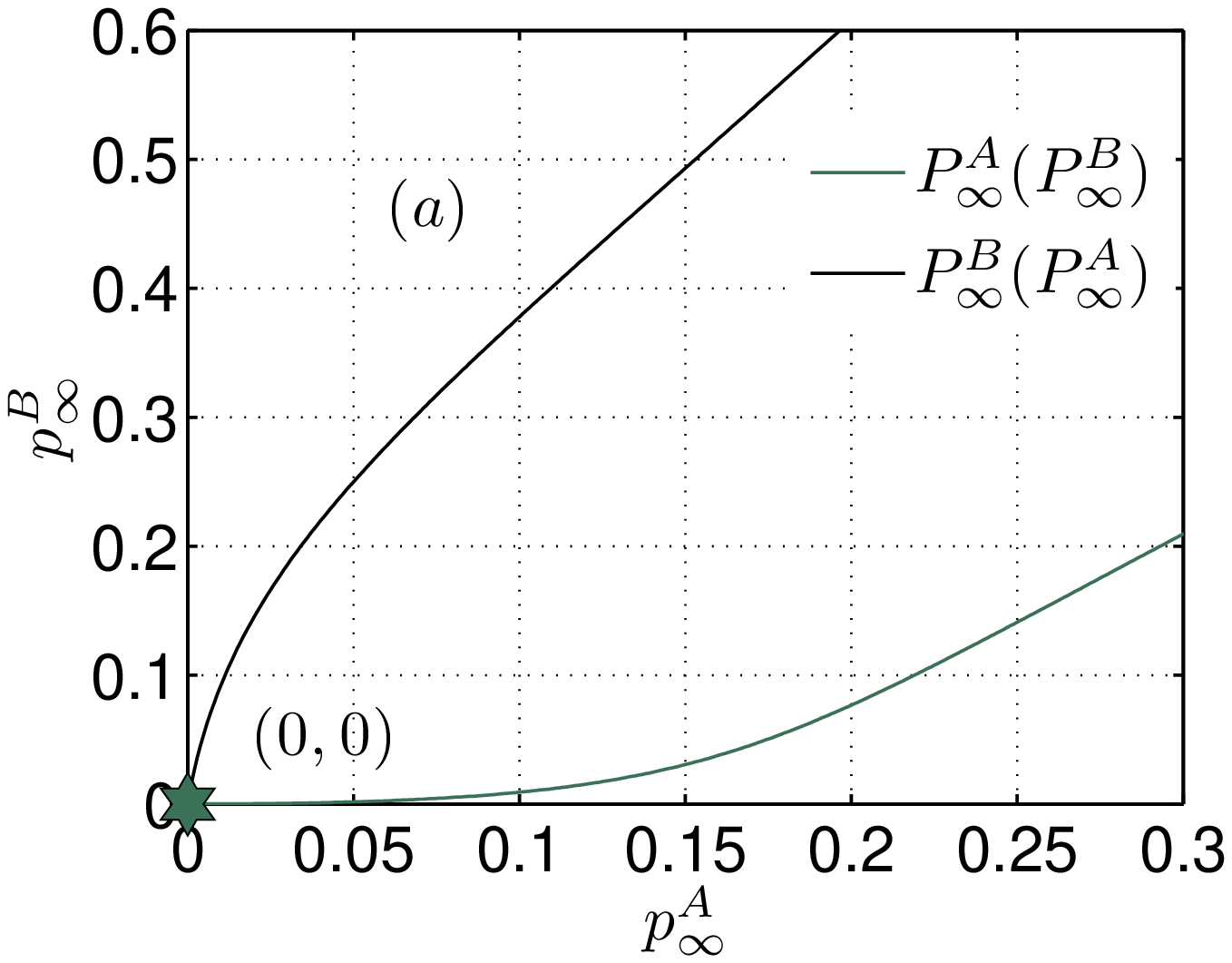}}
\centering \scalebox{0.35}[0.35]{\includegraphics{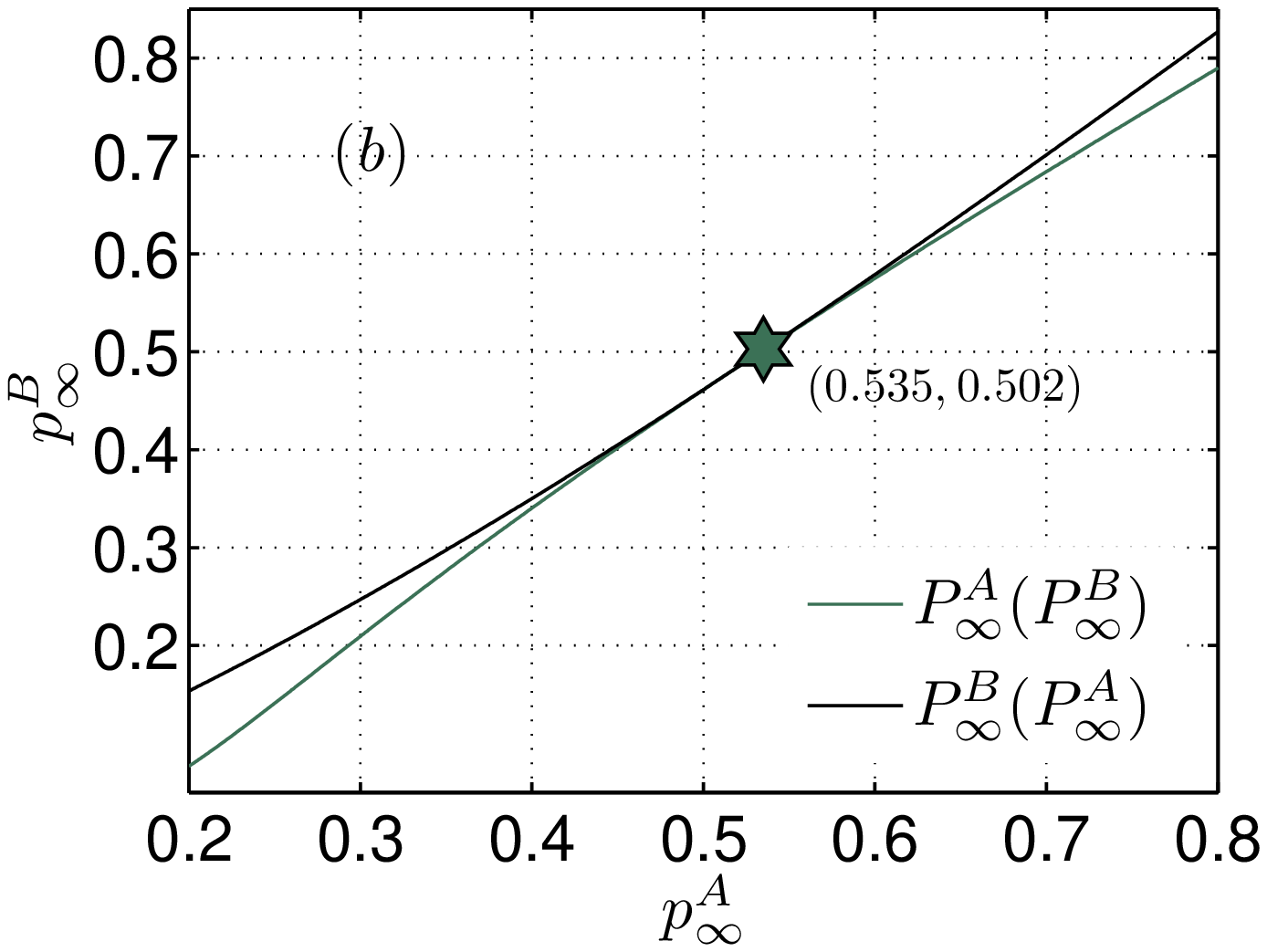}}
\centering \scalebox{0.35}[0.35]{\includegraphics{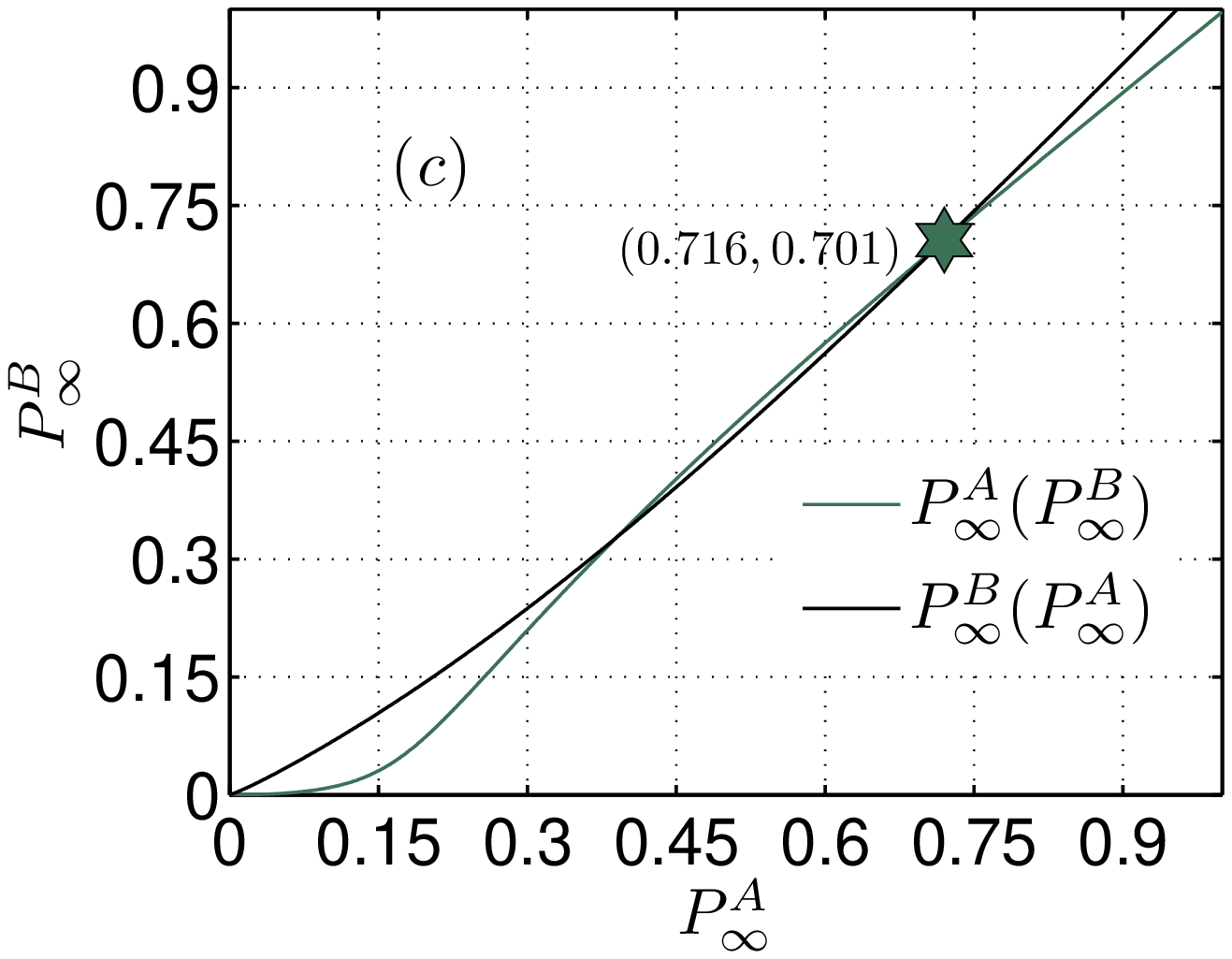}}
\caption{$P^{B}_{\infty}$ as a function of $P^{A}_{\infty}$ are
shown from Eq. (8) with the different $p$, $p=0.6$ (a), $p=0.94$
(b), $p=0.96$ (c). One can see that as $p<0.94$, only a trivial
solution $P^{A}_{\infty}=P^{B}_{\infty}=0$ exists from (a). As
$p=0.94$, the non-zero giant component of both networks appears at
stable state from (b). As $p=0.96>0.94$, the largest solution of two
curves is chosen, since the size of giant component gradually
decreases at cascading process.}
\end{figure}
From above analysis, the coupling strength $q_{A}$ as a function of
$p$ is studied from Eqs. (8) and (9), as shown in Fig. 5(a)-(c). We
can observe that as $q\in[0, q^{S,H}_{Ac}]$, the system only occurs
second order transition at $p^{II}_c$ from Fig. 5(a)-(c), where
coupling strength $q^{S,H}_{Ac}$ is a boundary point of between
second order region and hybrid order region. As $q\in(q^{S,H}_{Ac},
q^{H,F}_{Ac}]$, the system undergoes hybrid order transition and
have two critical points $p^{h,II}_{c}$ and $p^{h,I}_{c}$, where
coupling strength $q^{H,F}_{Ac}$ is a boundary point of between
hybrid order region and first order region. Similarly, as
$q>q^{H,F}_{Ac}$, the system only behaves first order transition and
$p^{I}_{c}$ appears. Furthermore, we can observe that when system
occurs second order transition behaviors, $p^{II}_{c}$ has a little
change as $\bar{k}$ increases from Fig. 5(d), which implies that
$\bar{k}$ has a little influence to robustness of system for weak
coupling. However, when system only undergoes first order transition
behaviors for strong coupling, $p^{I}_{c}$ decreases and system
become more robust as $\bar{k}$ increases. Especially, for coupling
strength $q_{A}$ corresponding hybrid order transition,
$p^{h,II}_{c}$ keep constant, $p^{h,I}_{c}$ decreases and eventually
coincidence, which suggests that hybrid order region disappears.
However, as $k$ increases, one can see that all the critical points
$p_{c}$ increases as $k$ increases from Fig. 5(d), which means the
system become robust as average degree of inter-network increases.
Fig. 6(a) and (b) describe that the phase transition region changes
as $\bar{k}$ and $k$ increase. We can see that first order region
gradually become larger due to $q^{H,F}_{Ac}$ increases as $\bar{k}$
increases from Fig. 6(a). Meanwhile, since the difference between
$q^{H,F}_{Ac}$ and $q^{S,H}_{Ac}$ become smaller, the hybrid order
region become smaller and eventually disappear as $\bar{k}$
increases. At this time, the system only occur first order
transition. Additionally, when $k$ increases, first order region
becomes smaller and
 second-order region becomes larger but hybrid order region almost keep constant.

\begin{figure}[H]
\centering \scalebox{0.5}[0.5]{\includegraphics{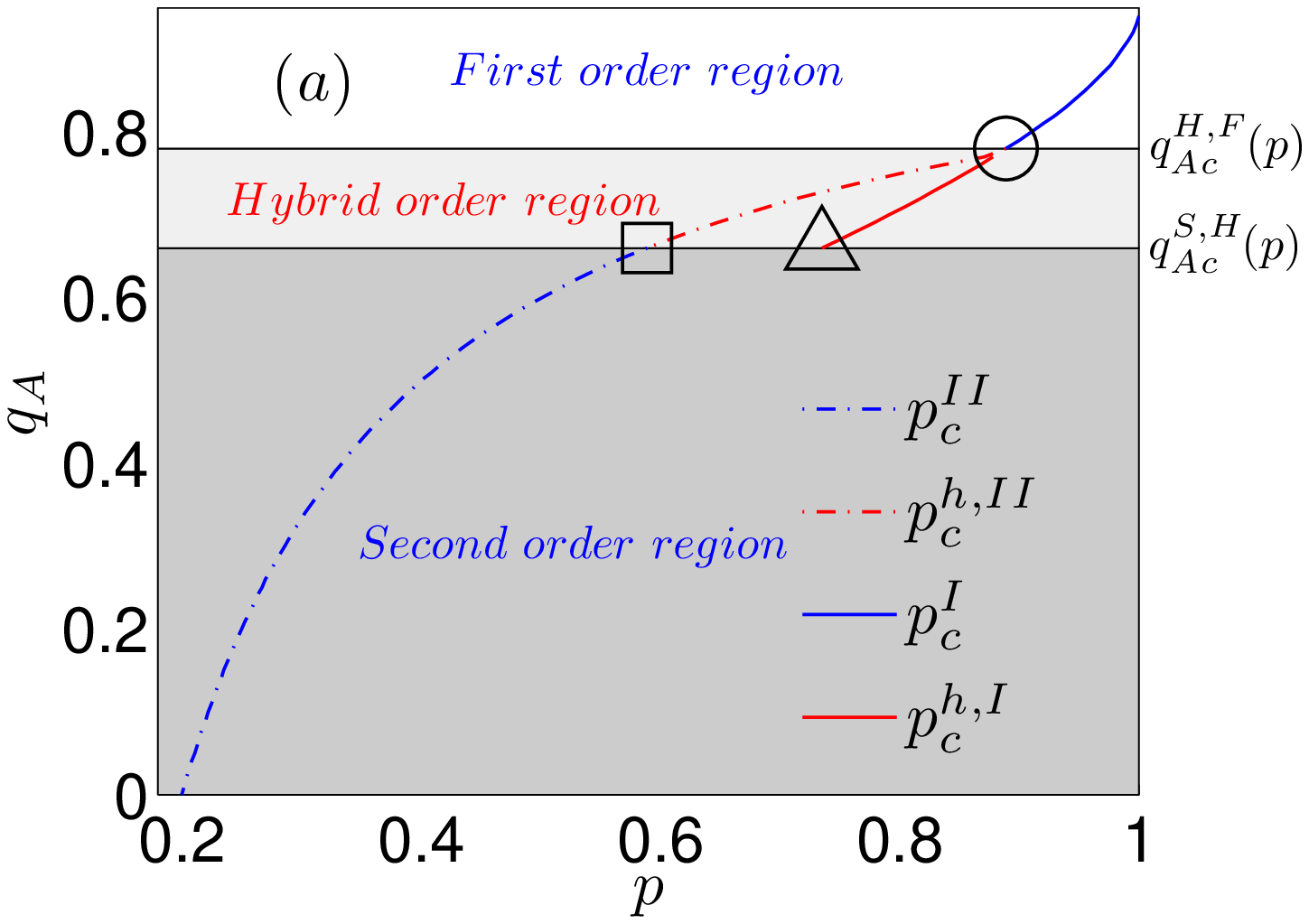}}
\centering \scalebox{0.5}[0.5]{\includegraphics{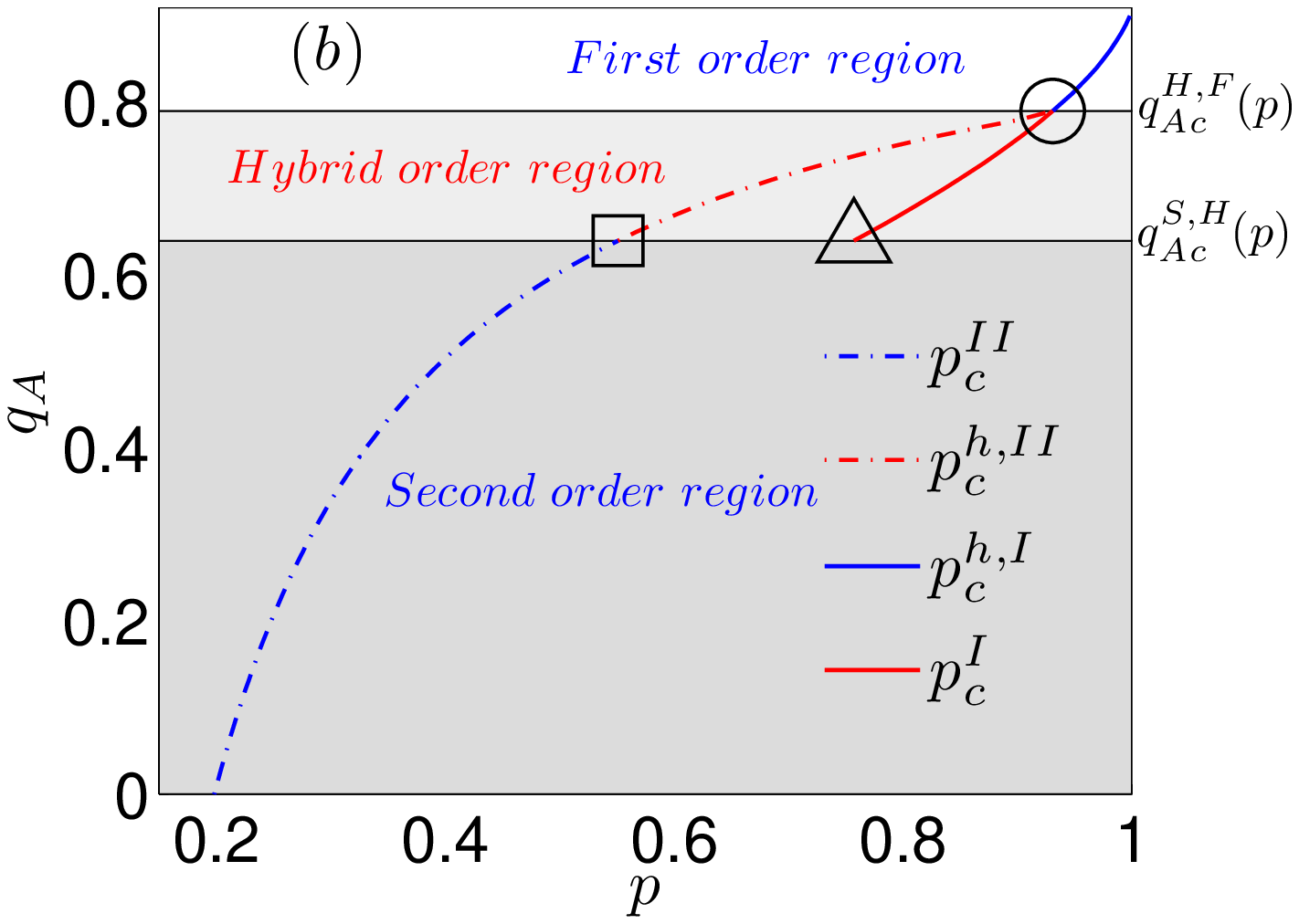}}
\centering \scalebox{0.5}[0.5]{\includegraphics{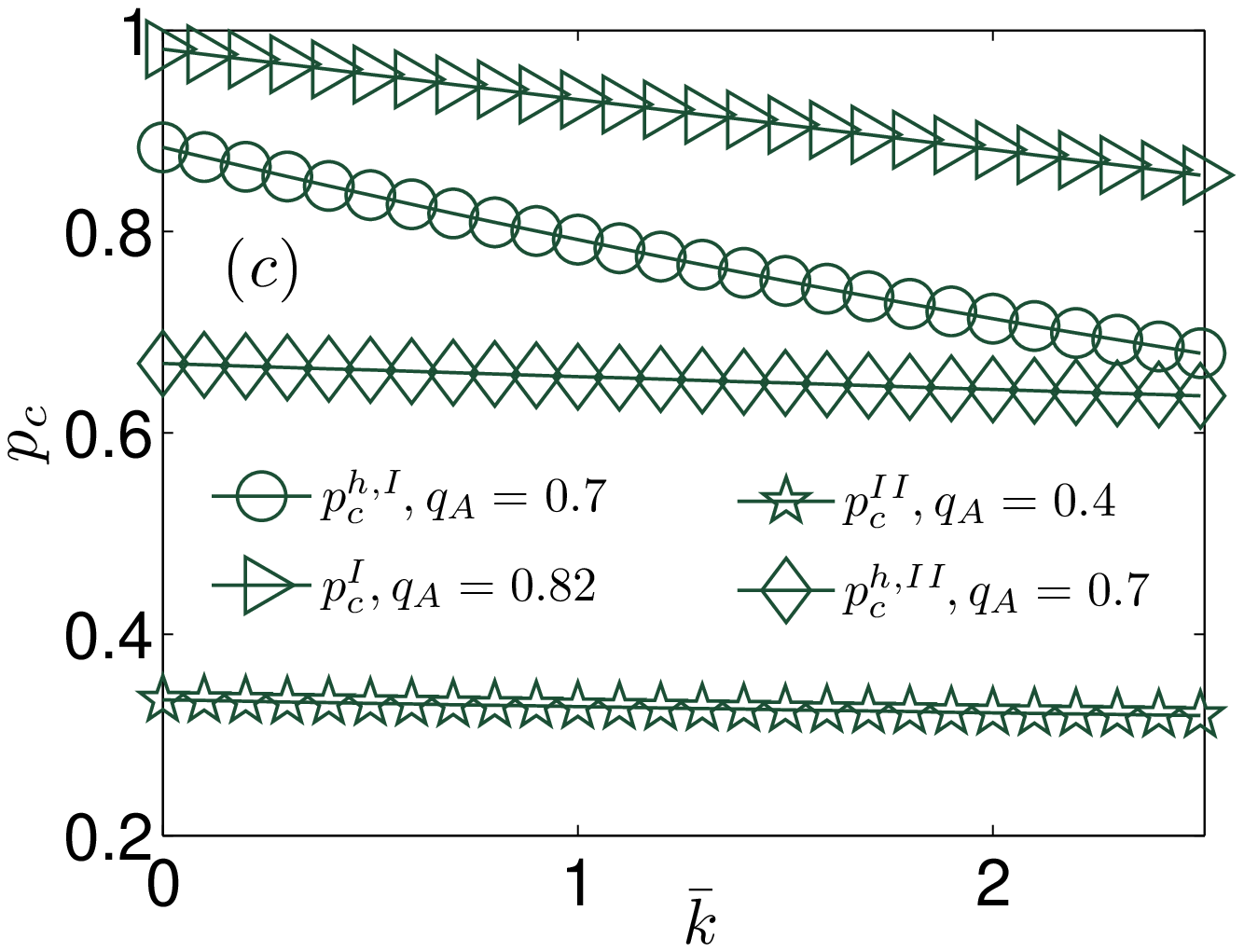}}
\centering \scalebox{0.5}[0.5]{\includegraphics{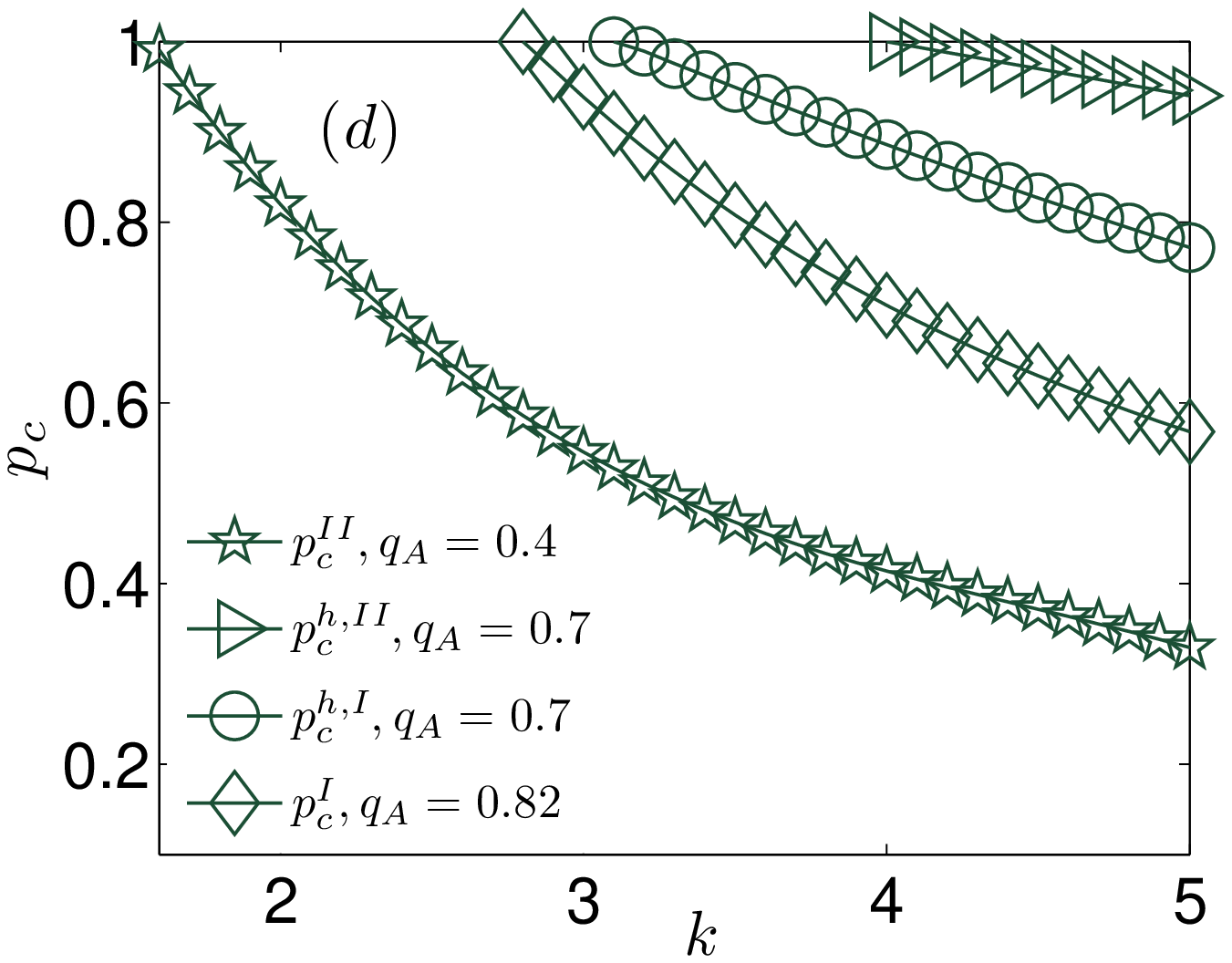}}
\caption{The coupling strength $q_{A}$ as a function of $p$ for
different parameter $\bar{k}$ with parameters $k_{A}=k_{B}=k=5$,
$\bar{k}_{A}=\bar{k}_{B}=\bar{k}$ and $q_{B}=1$. (a) $\bar{k}=1$.
(b) $\bar{k}=0.5$. (c) $p_c$ as a function of $\bar{k}$ for
different $q_{A}$ with parameters $k_{A}=k_{B}=k=5$ and $q_{B}=1$.
(d) $p_c$ as a function of $k$ for different $q_{A}$ with parameters
$\bar{k}_{A}=\bar{k}_{B}=\bar{k}=0.5$ and $q_{B}=1$.}
\end{figure}

\begin{figure}[H]
\centering \scalebox{0.5}[0.5]{\includegraphics{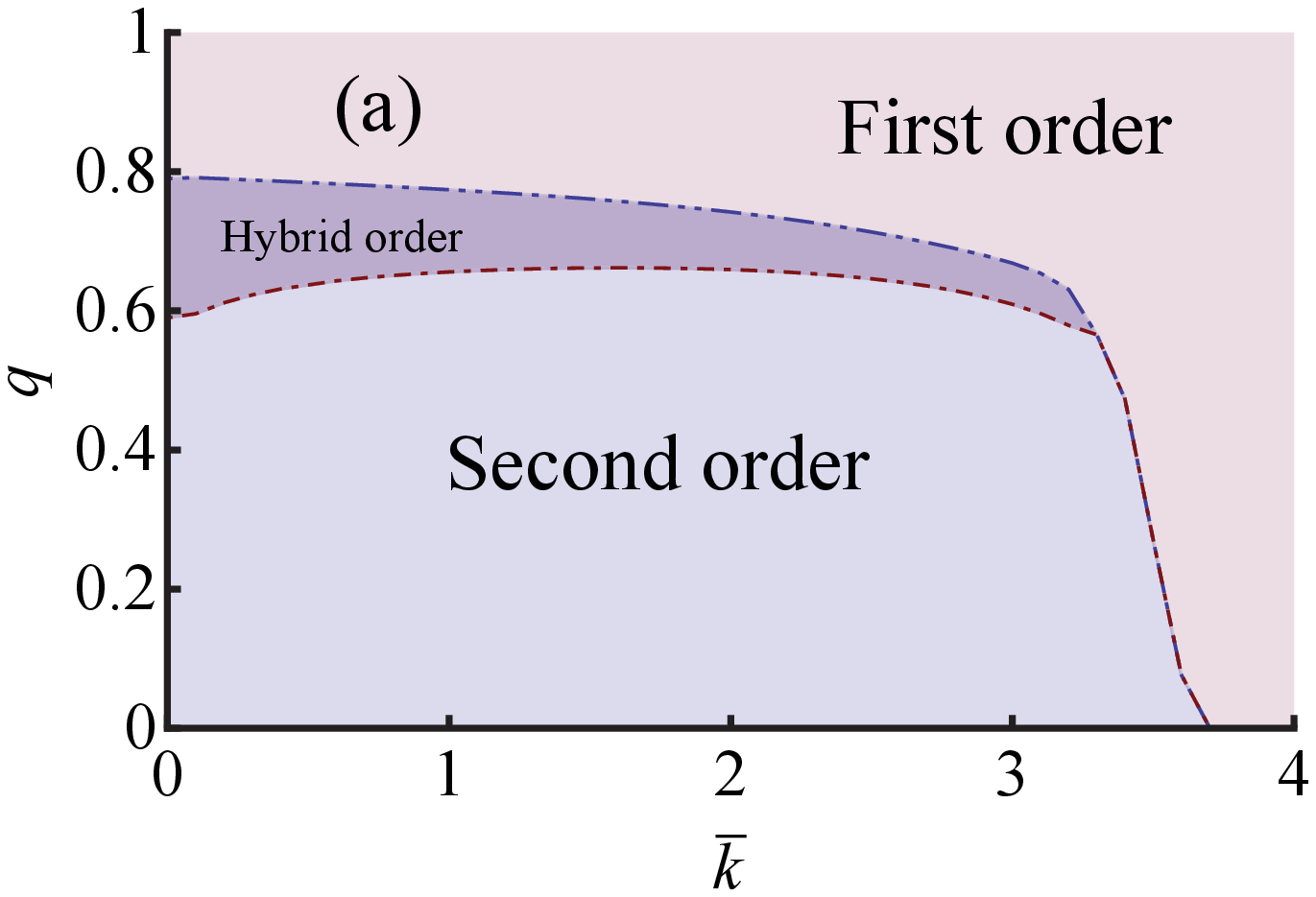}}
\centering \scalebox{0.5}[0.5]{\includegraphics{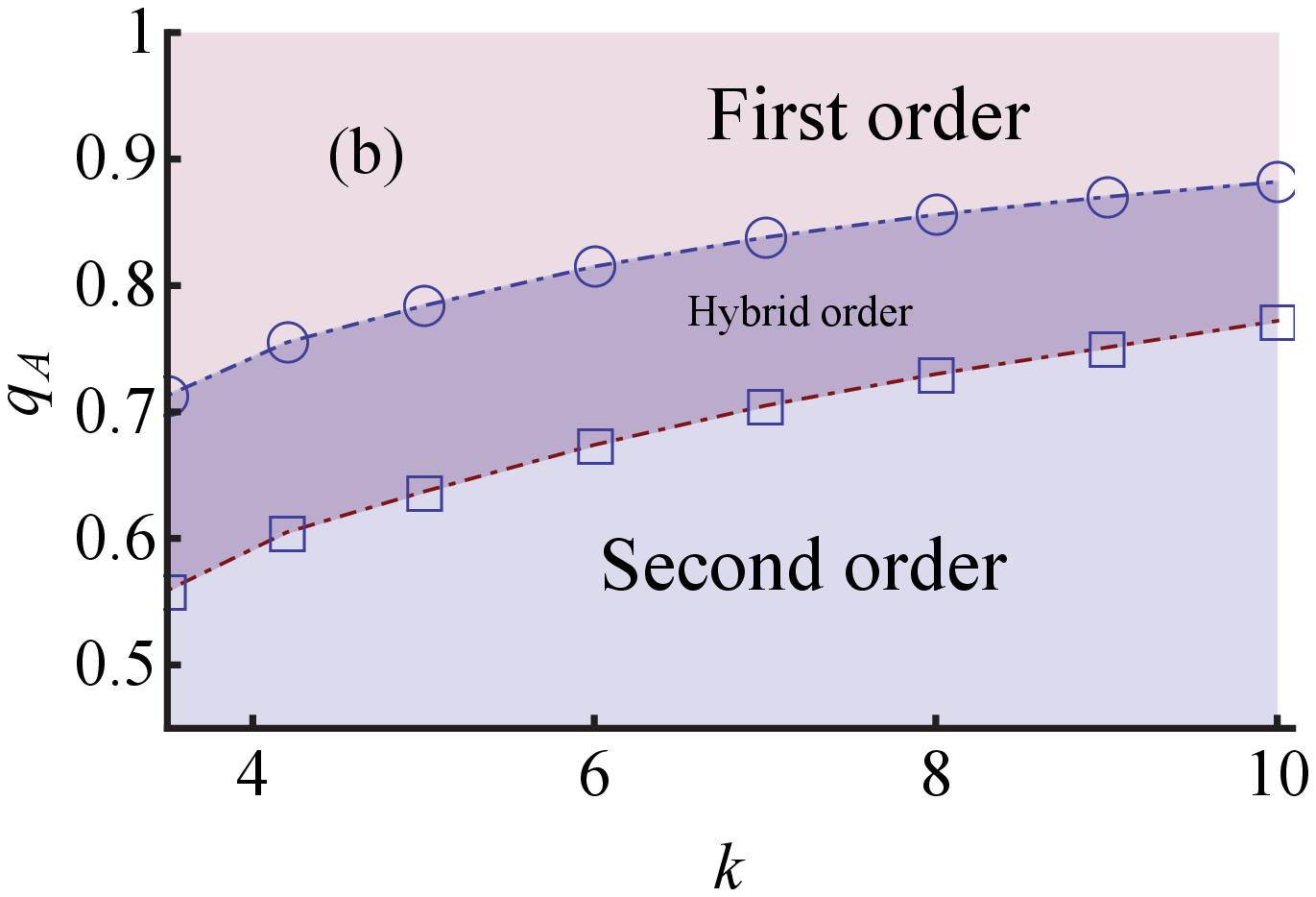}}
\caption{The coupling strength $q_{A}$ as a function of $\bar{k}$
(a) and $k$ (b) at the critical faction $p_{c}$. (a) the parameters
are $k_{A}=k_{B}=k=5$, $\bar{k}_{A}=\bar{k}_{B}=\bar{k}$ and
$q_{B}=1$. (b) The parameters are the similar with (a) but
$\bar{k}=0.5$. The blue and red dash line denote $q^{H,F}_{Ac}$ and
$q^{S,H}_{Ac}$ respectively. }
\end{figure}

Furthermore, we compare our model with model under non-feedback
condition for two interacting networks. For the same parameters, by
comparing Fig. 7(a) with Fig. 5(b), one can find that when
dependency links satisfy feedback condition, $p_{c}$ is more bigger
than that under non-feedback condition. Thus, for two coupling
links, feedback condition between two networks make the system
extremely vulnerable, which means that the system is difficult to
defend for feedback condition. And, for feedback condition,
$q^{H,F}_{Ac}$ and $q^{S,H}_{Ac}$ are smaller than that under
non-feedback condition, which means the system have bigger first
order region under randomly attacking as shown in Fig. 7(b).

\begin{figure}[H]
\centering \scalebox{0.5}[0.5]{\includegraphics{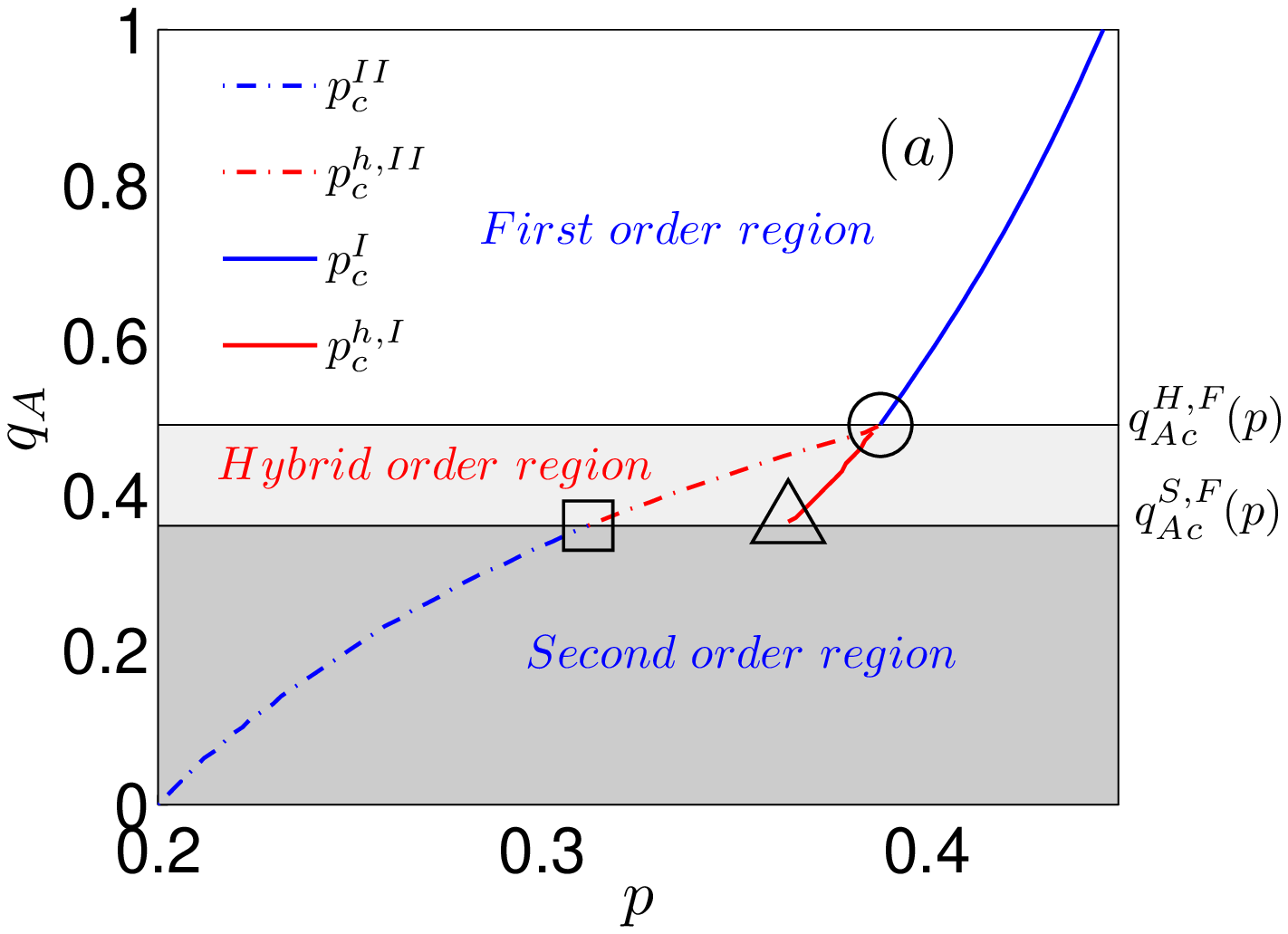}}
\centering \scalebox{0.55}[0.55]{\includegraphics{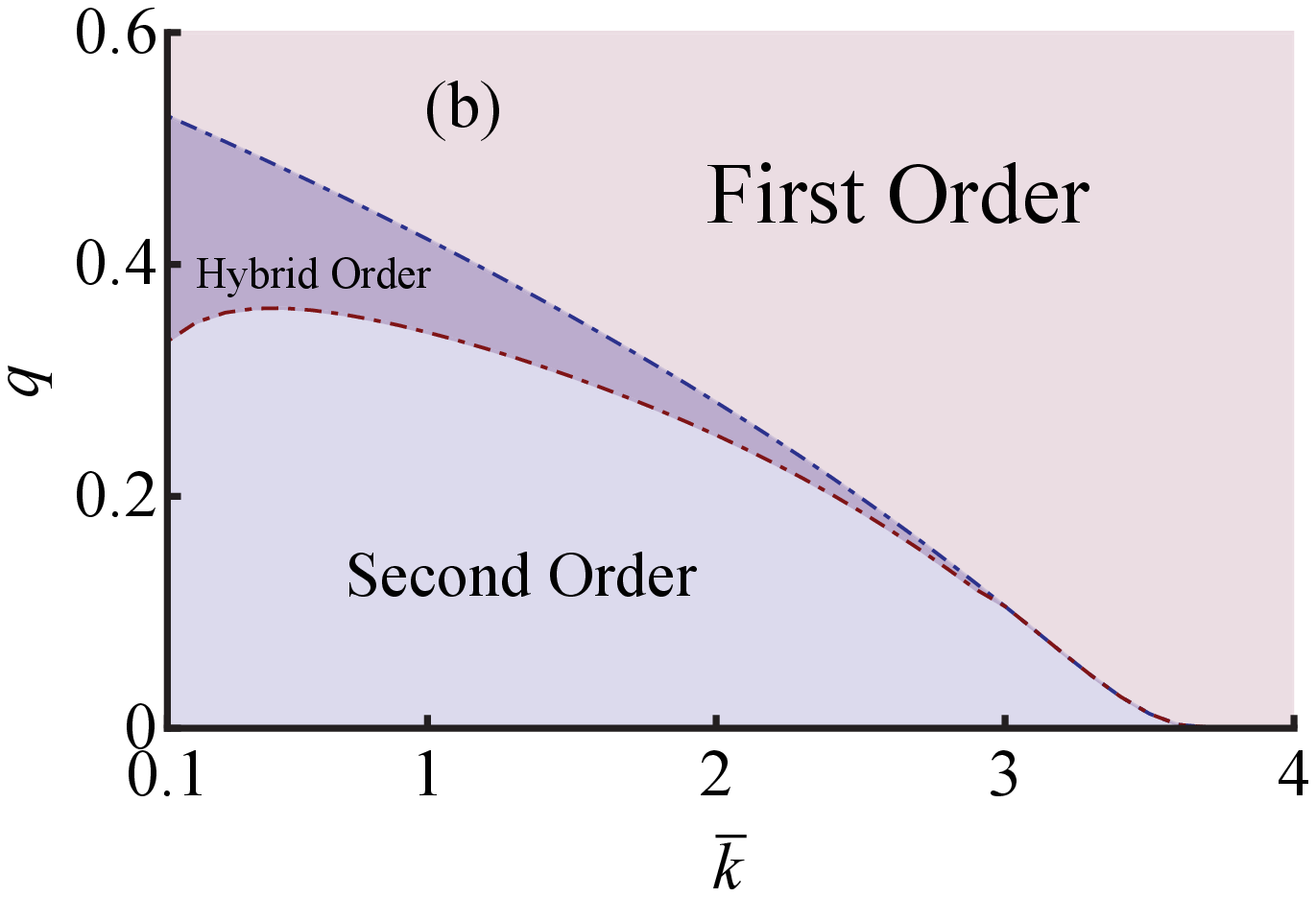}}
\caption{(a) The coupling strength $q_{A}$ as a function of
$\bar{k}$ at the critical faction $p_{c}$ under non-feedback
condition for the same parameters with Fig. 5(b). (b) The coupling
strength $q_{A}$ as a function of $p$ for different parameter
$\bar{k}$ for the same parameters with Fig. 6(a).}
\end{figure}

\section{conclusion}
In summary, we have introduced a framework for two interacting
network with feedback dependency links. Our theory is in excellent
agreement with the numerical simulations on coupled networks with
Poissonian distribution, which also can be applied to any degree
distribution networks. We find that for weak coupling strength,
$p^{II}_{c}$ has a little change and robustness of system is not
altered significantly as $\bar{k}$ increases. But for strong
coupling strength, $p^{I}_{c}$ decreases and the system become more
robust as $\bar{k}$ increases. However, for all the coupling
strength, the system become robust as $k$ increases. Moreover, as
$\bar{k}$ increases, $q^{S,H}_{Ac}$ and $q^{H,F}_{Ac}$ gradually
become small and eventually coincidence, which means that hybrid
order region disappears, and meanwhile the system only occurs first
and second phase transitions. Additionally, by comparing
non-feedback dependency condition between interacting networks, we
find that the system is extremely vulnerable and difficult to defend
for cascading failures.

\section{acknowledgments}

\end{document}